\documentstyle[aps,prb,eqsecnum,fleqn,preprint]{revtex}
\begin{document}
\draft
\title{The Superconducting Instabilities of the non half-filled Hubbard
Model in Two Dimensions}
\author{D. Zanchi and H. J. Schulz}
\address{Laboratoire de Physique des Solides, Universit\'e Paris-Sud, 91405
Orsay, France}
\maketitle
\begin{abstract}
The problem of weakly correlated electrons on a square lattice is formulated
in terms of one-loop renormalization group. Starting from the action for the
entire Brillouin zone (and not with a low-energy effective action) we reduce
successively the cutoff $\Lambda$ about the Fermi surface and follow the
renormalization of the coupling $U$ as a function of three
energy-momenta. We calculate the intrinsic scale $T_{co}$ where the
renormalization group flow crosses over from the regime ($\Lambda > T_{co}$)
where the electron-electron (e-e) and electron-hole (e-h) terms are equally
important to the regime ($\Lambda < T_{co}$) where only the e-e term plays a
role. In the low energy regime only the pairing interaction $V$ is
marginally relevant, containing contributions from all renormalization group
steps of the regime $\Lambda > T_{co}$. After diagonalization of $V_{\Lambda
=T_{co}}$, we identify its most attractive eigenvalue $\lambda _{\min}$. At
low filling, $\lambda _{\min}$ corresponds to the $B_2$ representation
($d_{xy}$ symmetry), while near half filling the strongest attraction occurs
in the $B_1$ representation ($d_{x^2-y^2}$ symmetry).  In the direction of
the van Hove singularities, the order parameter shows peaks with increasing
strength as one approaches half filling. Using the form of pairing and the
structure of the renormalization group equations in the low energy regime,
we give our interpretation of ARPES experiments trying to determine the
symmetry of the order parameter in the Bi2212 high-$T_{c}$ compound.
\end{abstract}

\bigskip


%\narrowtext

\section{Introduction}
An immense number of recent experiments on high $T_c$ superconductors aims
at a determination of the form of the BCS gap function in momentum
space.\cite{Schrieffer} Josephson junction
experiments,\cite{Wollman,Kirtley} measurements of the London penetration
depth,\cite{Dynes,Hardy} and of the Cu NMR relaxation rate \cite{Martindale}
are consistent with a $d_{x^2-y^2}$-gap. Particularly interesting are the
ARPES data which provide rather precise information about the detailed
angular dependence of the amplitude of the gap function. The experiments on
the Bi2212 compound show that the order parameter is maximal along the
$(0,\pi )$ direction \cite{Shen,Norman} and that its amplitude in the $(\pi
,\pi )$ direction seems to attain a nonzero value at a new critical
temperature below $T_c$. \cite{JianMa}

The $d$ symmetry of the gap function is generally considered to be a sign of
a pairing interaction of electronic origin, implying the absence of the
standard phononic mechanism for superconductivity.  The idea of a
superconducting state induced by fluctuations of purely electronic origin in
systems of electrons with Coulomb repulsion is originally due to Kohn and
Luttinger \cite{KohnLut} for the case of the three dimensional electron
gas. Similar effects exists in a two-dimensional electron gas, and generally
they depend strongly on the form of the Fermi surface. Perturbative
calculations of the four point vertex for a weakly filled band in the
Hubbard model show that the model is instable against $d_{xy}$
superconductivity \cite{Baranov}. Quantum Monte Carlo calculations on the
same model, in the vicinity of metal-insulator transition, show that the
attractive pairing interaction of $d_{x^2-y^2}$ symmetry is
dominant,\cite{Bulut}, in agreement with the earlier arguments that
antiferromagnetic fluctuations are the mediator of pairing interactions
\cite{Emery,Miyake,Scal1,Scal2,FLEX1,FLEX2,FLEX3}.
Direct evidence for $d_{x^2-y^2}$ superconducting
ordering has however not yet been found in quantum Monte Carlo
studies. Antiferromagnetic fluctuations become stronger and stronger as one
approaches half-filling. For weakly interacting electrons, these
fluctuations are associated with the $\ln ^2T$-divergence of the
electron--hole (e-h) loop diagram, caused by the nesting property of the
Fermi surface and the van Hove singularities. On the other hand the BCS
fluctuations, characterized by the electron-electron (e-e) loop, normally
only linear in logarithms, crosses over to a $\ln ^2T$ form in the vicinity
of half-filling. The perturbative treatment of an interacting system of
electrons should thus be based on the summation of all iterations of these
two types of loops.  The renormalization group is one way to do
this. Applied to interactions between electrons placed at the van Hove
points it gives an antiferromagnetic instability at half-filling and
superconductivity of $d_{x^2-y^2}$ symmetry if the deviation of the chemical
potential $\mu$ from its value at half-filling becomes of the order of
critical temperature of the antiferromagnetic state.\cite{HJS} The
equivalent parquet approach has been used for half-filling and also finds the
antiferromagnetic state.\cite{dzy,DzYak} A direct calculation of the zero
temperature free energy \cite{Georges} up to the second order in the bare
interaction $U_0$ confirms that at  half-filling the antiferromagnetic
order is stable, but finds no finite superconducting order parameter at any
filling.

In the present analysis we search to know whether the Hubbard model with
repulsive on-site interaction can lead to superconductivity, and if it does,
to what form of the gap function. In particular, we are interested in the
dependence of the results on the density of electrons. It is hoped that the
results can help us to clarify the origins of the existence of a highly
anisotropic BCS gap function in the cuprates.

The renormalization group technique for fermionic systems in two and three
dimensions has recently been developed, but with very drastic
limitations. The Wilsonian mode-elimination technique was applied by Shankar
\cite{Shankar} only to systems with an either isotropic or open, perfectly
nested Fermi surface. Weinberg \cite{Weinberg} has written the flow
equations for a general case of an anisotropic Fermi surface, but taking
into account only the electron-electron channel of the flow. Moreover, a
common tendency is to do the renormalization group procedure only for a thin
ring of degrees of freedom around the Fermi surface and to linearize the
spectrum in the radial direction, taking as a starting model the low-energy
effective action. This makes it difficult or impossible to make statements
about the phase diagram of lattice models like those relevant for the
description of the cuprates, due to the absence of a proper description of
high-energy degrees of freedom, the elimination of which may considerably
affect the effective low--energy action.

We formulate the renormalization group starting from the Hubbard model, with
the whole Brillouin zone involved in renormalization. We do no tree level
scaling before the energy shell $\pm \Lambda$ of available states around the
Fermi surface contains no van Hove singularity. Once $\Lambda$ has become
the smallest energy scale, performing a tree level analysis and deriving the
renormalization group equations, we show that the effective coupling
function contains some relevant contributions with the origins in the regime
with higher $\Lambda$, i.e. from the electronic degrees of freedom not
included in the low energy effective action. We must add that our
renormalization group, since perturbative in interaction, can provide
uniquely the information whether the Fermi liquid is a fixed point or
not. If the coupling flows to strong coupling, we can say in which direction
it flows, for example in the $d$--type superconducting direction, but we can
not say whether another fixed point with finite superconducting order
parameter with $d$ symmetry exists or not.  This kind of problem is well
known e.g. from the renormalization group in quasi-one-dimensional
compounds, where the most divergent flow in some direction is always
associated with the corresponding long--range order (LRO) because already
infinitesimal interchain coupling suffices for its
stabilization.\cite{HeinzCO} Similarly, the dimensionality reason for the
non-existence of LRO in two dimensions at finite temperature can be ignored
as soon as small hopping in the third direction exists. This, however,
doesn't mean that LRO and Fermi liquid fixed points are the only possible,
on this question the one-loop renormalization group simply can not give an
answer.

The problem is formulated in section 2 in terms of the effective action and
the corresponding renormalization group flow equation for the coupling
function. In section 3, after calculation of the crossover energy to the
purely electron-electron (e-e) part of the flow, we derive the
renormalization group equation for the pairing function $V$. In section
4. we diagonalize the pairing $V$ for the case of a non-divergent
electron-hole (e-h) channel and determine the most attractive eigenfunction
and the resulting critical temperature as functions of the chemical
potential $\mu$. In section 5 we give our picture of the gap viewed by ARPES
experiments.  The conclusions are given in section 6.

\section{Model and Formulation of the renormalization group theory}
The Hubbard model for a two-dimensional system of electrons on a square
lattice is given by
\begin{equation} \label{Hubbard}
H=\sum _{\sigma {\bf k}} \xi _{\bf k}^0
{a}^{\dagger}_{\sigma {\bf k}}
{a}_{\sigma {\bf k}}+
\frac{1}{2}U_0\sum _{\sigma }
\sum _{{\bf k}_1,{\bf k}_2,{\bf k}_3}
{a}^{\dagger}_{-\sigma,{\bf k_1+k_2-k_3}}
{a}_{-\sigma  {\bf k_2}}
{a}^{\dagger}_{\sigma {\bf k_3}}
{a}_{\sigma {\bf k_1}}\; ,
\end{equation}
where $\xi _{\bf k}^0=-2t(\cos k_x + \cos k_y)-\mu$ and the momenta are
within the first Brillouin zone.  For this model we introduce now the action
in terms of fermion coherent states represented by Grassmann
variables\cite{Negele} ${\Psi}_{\sigma K}$ and $\bar{\Psi}_{\sigma K}$,
where $K=({\bf k},\omega )$. We will write it in the form
$$
S\{ \Lambda _0,\xi _{\bf k}^0,U_0 \} =\int _{0}^{\infty}\frac{d\tau}{2\pi}
\sum _{\sigma
{\bf k}} \Theta (\Lambda _0-|\xi _{\bf k}^0|) \bar{\Psi}_{\sigma K} (\partial
_{\tau}
-\xi _{\bf k}^0) {\Psi}_{\sigma K}+
$$
\begin{equation} \label{act}
+\frac{1}{2}\sum _{\sigma \sigma '}\int \left(\prod _{i=1}^3
\frac{d\omega _i}{2\pi}\right)
\sum
_{\bf k_1,k_2,k_3}
U_0\Theta ^{(\Lambda _0)}_{\bf
k_1,k_2,k_3,k_4}
\bar{\Psi}_{\sigma K_3}\bar{\Psi}_{\sigma 'K_4}{\Psi}_{\sigma 'K_2}
{\Psi}_{\sigma K_1}
\; ,
\end{equation}
where $\Theta ^{(\Lambda )}_{\bf k_1,k_2,k_3,k_4}\equiv \prod
_{i=1}^{4}\Theta (\Lambda-|\xi _{\bf k_i}^0|)$ constrains all four momenta
to run within the energy shell $\pm \Lambda _0=8t$ around the Fermi surface.
The energy and momentum are conserved so that $K_4(K_1,K_2,K_3)=(\omega
_1+\omega _2-\omega _3, {\bf k_1+ k_2 - k_3})$. Note that the size of the
cutoff is equal to the bandwidth, i.e. the whole Brillouin zone is
available for integration. Thus, the $\Theta $ functions have no meaning
yet: they become important when the cutoff, reduced by the renormalization
group, become lower than a tilt from the Fermi level to the band
boundary. Note that for a non-half filled band the effective phase space is
not particle-hole symmetric.

The renormalization group transformation that we will use, known as the
field theory approach, is defined as the mapping
\begin{equation} \label{RG}
S\{ \Lambda _0,\xi _{\bf k}^0,U_0 \}\rightarrow
S'=S\{\Lambda _0 \rightarrow \Lambda _0 e^{-l},\xi _{\bf k}^0\rightarrow \xi
_{\bf k},U_0\rightarrow U(K_1,K_2,K_3) \} ,
\end{equation}
where $\xi_{\bf k}$ and $U$ depend on $l$ in a way so that the physical
properties of $S'$ and $S$ are the same for energies lower than $\Lambda
=\Lambda _0 e^{-l}$.  This requirement is fulfilled if all one-particle
irreducible vertices are invariant under reduction of the cutoff from
$\Lambda _0$ to $\Lambda _0 e^{-l}$.  The renormalization group can be
thought of as a set of successive, infinitesimally small steps $dl$.  This
allows us to formalize the renormalization group requirement in a set of
equations $\partial _l\Gamma _i=0$, where $i=2,4,6,...$.  Up to second order
in $U$, it suffices to consider only $\Gamma _2$ and $\Gamma _4$, because
all higher vertices are of higher order in $U$.  The solutions of the
equations $\partial _l\Gamma _2=0$ and $\partial _l\Gamma _2=0$ give us the
renormalization group flow for $\xi _{\bf k}$ and $U(K_1,K_2,K_3)$.

Conservation of spin allows us to write the interaction part of the action
as a sum of the singlet $(|\vec{\sigma }+\vec{\sigma '}|=0)$ and triplet
$(|\vec{ \sigma }+\vec{\sigma '}|=\sqrt{2})$ parts:
\begin{equation} \label{SingTrip}
\bar{s}(K_4,K_3)U^S(K_1,K_2,K_3)s(K_2,K_1)+\bar{t}_{\mu}
(K_4,K_3)U^A(K_1,K_2,K_3)t_{\mu}(K_2,K_1),
\end{equation}
where $s$ and $t_{\mu}$ are the variables of annihilation of the
singlet and triplet states
 \begin{equation} \label{Sing}
s(K_2,K_1)\equiv \frac{1}{\sqrt{2}}\sum _{\sigma}\sigma{\Psi}_{\sigma K_2}
{\Psi}_{-\sigma K_1},
\end{equation}
\begin{equation} \label{Trip}
t_{0}(K_2,K_1)\equiv \frac{1}{\sqrt{2}}\sum _{\sigma }{\Psi}_{\sigma K_2}
{\Psi}_{-\sigma K_1}\hspace{10mm} ; \hspace{10mm} t_{\pm 1}(K_2,K_1)\equiv
{\Psi}_{\uparrow ,\downarrow K_2}
{\Psi}_{\uparrow ,\downarrow  K_1}.
\end{equation}
The singlet state is symmetric and the triplet antisymmetric under exchange
of the momenta of two particles. Correspondingly, the coupling function
$U^S(K_1,K_2,K_3)$ can be taken to be symmetric and $U^A(K_1,K_2,K_3)$ to be
antisymmetric under the momentum exchange operation $X$, defined as
\begin{equation} \label{X}
X{\cal F}(K_1,K_2,K_3)={\cal F}(K_2,K_1,K_3)\; ,
\end{equation}
${\cal F}$ being a function of four energy--momenta which conserves energy
and momentum.  If ${\cal F}$ possesses time-reversal symmetry
\begin{equation} \label{Timerev}
{\cal F}(K_1,K_2,K_3)={\cal T}{\cal F}(K_1,K_2,K_3)\equiv {\cal
F}(K_3,K_4(K_1,K_2,K_3),K_1),
\end{equation}
which certainly is a property of the vertex, then it is equivalent whether
$X$ exchanges $K_1$ and $K_2$ or $K_3$ and $K_4$, i.e. ${\cal
F}(K_2,K_1,K_3)={\cal F}(K_1,K_2,K_4(K_1,K_2,K_3))$.  Formally, $U^S$ and
$U^A$ are given by
\begin{equation} \label{AS}
U^A=\frac{1}{2}(1-X)U\; ,\hspace{10mm}
U^S=\frac{1}{2}(1+X)U\; .
\end{equation}
On the other hand, the interaction can also be written as a sum of one term
with equal $(\sigma =\sigma ')$ and one with opposite $(\sigma =-\sigma ')$
spin quantum numbers, with corresponding coupling functions named
$U_{\parallel}(K_1,K_2,K_3)$ and $U_{\perp}(K_1,K_2,K_3)$, respectively. From
two equal-spin electrons one can build only a triplet state, which make us
conclude that
\begin{equation} \label{Upar}
U_{\parallel}=U^A,
\end{equation}
while
\begin{equation} \label{Uper}
 U_{\perp}=U=U^A+U^S\; ,
\end{equation}
containing the singlet and the triplet interactions.

Another way to write the interaction is in terms  of a charge and a spin part
\begin{equation} \label{SpinCharge}
U_c(K_1,K_2,K_3)\bar{C}(K_2,K_4)C(K_3,K_1)+U_{\sigma}(K_1,K_2,K_3)
\bar{\bf S}(K_2,K_4)\cdot {\bf S}(K_3,K_1)\; ,
\end{equation}
where $C$ and $S_i$ are
\begin{equation} \label{CSoper}
C(K_3,K_1)\equiv \sum _{\sigma} \bar{\Psi} _{\sigma K_3}\Psi _{\sigma
K_1} \hspace{10mm};\hspace{10mm} S_i(K_3,K_1)=\sum _{\sigma \sigma '}\bar{\Psi}
_{\sigma K_3} \sigma ^i_{\sigma \sigma '} \Psi _{\sigma '
K_1}\; .
\end{equation}
The charge and spin coupling functions are
\begin{equation} \label{CSinter}
U_c=\frac{1}{4}(2-X)U \; , \hspace{10mm} U_{\sigma}=-\frac{X}{4}U\; .
\end{equation}
In first order of perturbation theory, these functions determine the
renormalization of the charge-charge and the spin-spin correlation
functions.

We proceed now with the derivation of renormalization group equations. For
simplicity, we will ignore the renormalization flow of $\xi _{\bf k}$, which
follows from conservation of $\Gamma _2$, renormalizing the form of the
Fermi surface, the effective mass, etc. This approximation is justified in
the case of the circular Fermi surface \cite{Shankar}. In the anisotropic
case, the diagrams for $\Gamma _2$ have a dependence on the direction of
{\bf k}. Moreover, even a small renormalization of the Fermi energy can give
important changes of the form of the Fermi surface if one is close to
half--filling, because of van Hove singularities.  For filling not too close
to one-half we can expect that the essential of the physics is given by just
the renormalization of the coupling $U$ using the bare dispersion relation
$\xi _{\bf k}^0$, which we will call $\xi _{\bf k}$ from now on.

The Feynman diagrams for $\Gamma _2$ and $\Gamma_{4 \perp}=\Gamma _{4}$ are
given in Fig. 1.  The first loop in the expression for $\Gamma _4$ is of the
electron-electron (e-e) and all others of the electron-hole (e-h) type.
Making use of the relations (\ref{Upar}), (\ref{Uper}), and (\ref{AS}), we
get the expression for $\Gamma _4$ in terms of $U$ and $XU$.  If we write
the integration measure of the loop diagrams in the form
\begin{equation} \label{measure}
\int \frac{d\omega}{2\pi} \int _{-\Lambda}^{+\Lambda}d\xi \oint
\frac{ds}{v(s,\xi )}\; ,
\end{equation}
$s$ being the curves of constant energy $\xi$, then $d\Gamma _{4}$
corresponds to the integration of the two energy shells of width $|\Lambda
|dl$ at $\xi =\pm \Lambda$.  We obtain the following flow equation
\begin{equation}
\frac{\partial{U}}{\partial{l}}=\beta _{ee}\{ U,
U\} +\tilde{\beta}_{eh}\{
U,U\}\; ,
\label{flowU}
\end{equation}
with
\begin{equation} \label{Betatilde}
\tilde{\beta}_{eh}\{ U,U\}=2{\beta}_{eh}\{ U,U\}-
{\beta}_{eh}\{ U,XU\}-{\beta}_{eh}\{ XU,U\}-X{\beta}_{eh}\{ XU,XU\}
\end{equation}
The functionals $\beta _{ee}\{ U_1,U_2\} $ and ${\beta} _{eh}\{ U_1,U_2\} $
are the partial derivatives with respect to $l$ of the e-e and e-h loops and
both are bilinear forms in $U_1(K_1,K_2,K_3)$ and $U_2(K_1,K_2,K_3)$. They
read
\begin{equation} \label{Betaee}
\beta _{ee}\{ U_1,U_2\}=\left( \Xi \{ U_1,U_2\} +\Xi \{ XU_1,XU_2\} \right)
\frac{1+\kappa _{ee}}{2}
\end{equation}
and
\begin{equation} \label{Betaeh}
\beta _{eh}\{ U_1,U_2\}=\left( \Pi \{ U_1,U_2\} +{\cal T}\Pi \{
U_1,U_2\}\right)
\frac{1+\kappa _{eh}}{2},
\end{equation}
with
$$
\Xi \{ U_1,U_2\} =\frac{-\Lambda}{(2\pi)^2}\sum _{\nu =+,-}
\int \frac{ds_{\nu}}{v_{\nu}}
\Theta
\left(\Lambda
- |\xi _{{\bf k}_{\nu}-{\bf q}_{ee}}|\right)
\int _{-\infty}^{+\infty} \frac{d \omega}{2\pi}\;
\frac{1}{i\omega -\nu \Lambda}\;
\frac{1}{i(-\omega +\omega _{ee})-
\xi _{{\bf k}_{\nu}-{\bf q}_{ee}}}\times
$$
\begin{equation} \label{Xi}
\times  U_1(K_1,K_2,K_{(\nu )})U_2(K_3,K_4,K_{(\nu )})\; ,
\end{equation}
$$
\Pi \{ U_1,U_2\} =\frac{-\Lambda}{(2\pi)^2}\sum _{\nu =+,-}
\int \frac{ds_{\nu}}{v_{\nu}}
\Theta
\left(\Lambda
- |\xi _{{\bf k}_{\nu}+{\bf q}_{eh}}|\right)
\int _{-\infty}^{+\infty} \frac{d \omega}{2\pi}\;
\frac{1}{i\omega -\nu \Lambda} \;
\frac{1}{i(\omega +\omega _{eh})-
\xi _{{\bf k}_{\nu}+{\bf q}_{eh}}}\times
$$
\begin{equation} \label{Pi}
\times U_1(K_1,K_{(\nu )},K_3)U_2(K_4,K_{(\nu )},K_2)\; .
\end{equation}
The index $\nu =+,-$ symbolizes two energy shells at $+ \Lambda$ and
$-\Lambda$; $v_{\nu}$ stands for $v(s_{\nu},\xi =\nu \Lambda)$; $\omega
_{ee}\equiv \omega _1+\omega _2$; $\omega _{eh}\equiv \omega _1-\omega _3$;
${\bf q}_{ee} \equiv {\bf k}_1+{\bf k}_2$; ${\bf q}_{eh} \equiv {\bf
k}_1-{\bf k}_3$; $K_{\nu}\equiv ({\bf k}_{\nu},\omega) $, where $ {\bf
k}_{\nu}$ is the momentum running along the path $s_{\nu}$. $\kappa
_{ee}$ and $\kappa _{eh}$ are non-analytic functions of the
momenta, given by
$$
\kappa _{ee}= \left\{
\begin{array}{l}
	  0 \hspace{10mm} \mbox{for} \hspace{3mm} {\bf q}_{ee}=0 \\
        1\hspace{10mm} \mbox{otherwise}
\end{array} \right.
\hspace{5mm};\hspace{5mm}
\kappa _{eh}= \left\{
\begin{array}{l}
	  0 \hspace{10mm} \mbox{for} \hspace{3mm} {\bf q}_{eh}=(\pm \pi,\pm
\pi)  \\
        1\hspace{10mm} \mbox{otherwise}
\end{array} \right.
$$
Their origin is in the derivatives over  $\Lambda$
of the products like
$$ \Theta (\Lambda-|\xi _{\bf k+q}|)\Theta (\Lambda- |\xi _{\bf k}|)$$
when $\xi _{\bf k}=\xi _{\bf k+q}$.

\section{The temperature scale $T_{co}$}
A particularity of the renormalization group approach treating e-e and e-h
fluctuations in more than one dimension is the absence of self-similarity of
the problem. In fact, there is an intrinsic energy scale which is a function
of the band filling. It is associated with charge and spin fluctuations
coming from the e-h term.  We will proceed by estimating the characteristic
energy scales which appear in $\beta_{ee}$ and $\beta _{eh}$, when all four
particles are at the Fermi surface, with zero energy. If we are exactly at
half-filling, it is known that in the limit $\omega \rightarrow 0$ both e-e
and e-h loops scale like $l^2$, which corresponds to the square-logarithmic
divergence in both channels. This gives an explicit $\sim l$ dependence in
the $\beta $-functionals. Let us suppose now that the filling is slightly
lower than one-half, i.e. that $\mu $ is small and negative. We expect two
regimes. One is for $l\lesssim l_x \sim \ln |8t/\mu |$, where the flow is still
unaffected by the small changes of the Fermi surface due to nonzero $\mu $
and remains proportional to $l$. In the second regime, where $l\gtrsim l_x$,
the e-e flow is just a constant (i.e. only a $\ln$-divergence), while the
e-h flow decays exponentially due to disappearance of nesting. Even far from
half-filling it is possible to define a crossover $l_x$, beyond which the
flow in the e-h channel disappears exponentially. We can summarize saying
that for any filling, $l_x$ is a crossover from a regime where both e--e and
e--h loops contribute to a regime where $\beta _{eh}$ starts to behave like
$\beta _{eh}\sim \Lambda ^{\eta ({\bf q})}$. Here $ \eta ({\bf q})$ is
positive for all values of the momentum transfer ${\bf q}={\bf q}_{eh}$.

To estimate the dependence of $l_x$ on the filling $\langle n\rangle $, we
consider the static limit of the partially integrated e-h loop
\begin{equation} \label{Bubble}
P_{eh}(l,{\bf q},\omega =0)=\frac{1}{U_0^2}\int_0^{l}\beta _{eh}\{ U_0, U_0\}.
\end{equation}
with the momentum transfer {\bf q} equal to $2{\bf k}_F$ in the direction
$(\pi, \pi )$.  Note that the energy integration is performed over $8t<\xi
<8t\exp (-l)$.  The derivative of $P_{eh}(l,{\bf q},\omega =0)$ with respect
to $l$ gives the explicit $l$-dependence in the $\beta _{ee}$
functional. Fig. 2 shows $P_{eh}(l)$ and $\partial _lP_{eh}(l)$ for two
different values of $\mu$.  It is reasonable to define $l_x$ as the point
where $\partial _lP_{eh}(l)$ starts to decrease.  In the exponential regime
the function $\partial _lP_{eh}(l)$ decays like $\exp (-l/2)$ (i.e. $\eta
(2{\bf k}_F)=1/2$, valid for any orientation of ${\bf k}_F$), while the
regime $l<l_x$ remembers the $\ln ^2$ divergence of $P_{eh}$ at half
filling. If we consider $P_{eh}(l)$ for some large momentum transfer
different from $2{\bf k}_F$ (giving intersection rather than touching of the
initial- and final-state Fermi surface) we get a shape like $\exp (-l)$
(i.e. $\eta ({\bf q}\neq 2{\bf k}_F )=1$).  The dependence $l_x(\langle
n\rangle )$ is shown in Fig.3. Near $\langle n\rangle =1$ there is a
divergence of the form $l_x(\langle
n\rangle )\approx \ln |8t/\mu (\langle n\rangle )|$ because of nesting,
while the increasing $l_x$ as the filling goes to zero mirrors the fact
that, for low density, the Fermi energy appears as the new scale instead of
the band-width being used.  The insert shows the function $\langle n\rangle
(\mu )$.

Once in the exponential regime, $\beta _{eh}$ can be neglected after it
becomes smaller than $e^{-1}$ of its value at $l=l_x$. Putting $\eta =1/2$,
this defines the crossover
\begin{equation}
\label{co}
l_{co}(\mu )=l_x(\mu )+2\; ,
\end{equation}
corresponding to the crossover temperature $T_{co}=8t\exp (-l_{co})$.
Suppose that we now integrate the flow equation (\ref{flowU}) from $l=0$ to
$l=l_{co}$. Once $l$ has reached $l_{co}$, only the term $\beta_{ee}\{ U\} $
remains in the flow equations, and one has a partially renormalized
$U(l=l_{co})$ as initial condition.  Now we use the fact that $T_{co}/8t$ is
a small parameter, i.e. the inequality $|\xi _{\bf k}|<T_{co}$ determines a
thin ring of degrees of freedom, containing no van Hove points, as one can
conclude looking at Fig.3.  This allows us to rescale the momenta
$k_{\perp}={\bf\hat{n}}({\bf k-k_F})$, where ${\bf\hat{n}}$ is the unit
vector normal to Fermi surface, dependent on direction of {\bf k}.  To
clarify the reason for which a tree-level scaling is not allowed for
energies higher than the deviation of the Fermi level from van Hove
singularity, let us write the phase space integration measure in terms of
energy $(\xi )$ and polar angle $(\theta )$ variables
\begin{equation} \label{measure0}
\frac{1}{2\pi}\int d{\bf k}=\frac{1}{2\pi}\int d\xi d\theta  J(\xi,\theta)
\end{equation}
with $J(\xi,\theta)= k(\xi,\theta)/v(\xi,\theta)$, $k$ being the radial wave
number and $v$ the group-velocity. Zeroth order (tree level) scaling tells
us via power counting argument to consider $J(\xi,\theta)$ as function of
$\theta$, neglecting any $\xi$ dependence about $\xi=0$, which is possible
if $J(\xi,\theta)$ is an analytic function of $\xi$ at the whole shell $\pm
\Lambda$ which, consequently, should contain no singularity.

We can also rescale the frequencies if $U(l_{co})$ is an analytic function
of $\omega $ in the interval $\pm T_{co}$ about the Fermi surface, which we
assume to be the case\cite{DupuisFL}.  In the scope of this tree-level
scaling, as it has already been shown by Shankar \cite{Shankar}, the slope
of the electronic dispersion around the Fermi surface is irrelevant and the
two marginal interactions correspond to two different constraints on
the four-momenta in $U$. Since any $k_{\perp}$- and $\omega$-dependence in $U$
is irrelevant, both marginal interactions depend only on coordinates of the
zero frequency particles placed at the Fermi surface. For the first, ``Fermi
liquid'' or forward interaction, the momenta satisfy the equation ${\bf
k}_1={\bf k}_3$, where the meaning of momenta can be seen from the equation
(\ref{act}).  This interaction is slightly $(\sim U_0^2)$ renormalized by
the high-energy-modes ($l<l_{co}$), and is not involved in further
renormalization.  The second interaction is the pairing potential $V$, where
the momenta satisfy the condition ${\bf k}_1=-{\bf k}_2$.  The pairing $V$
depends only on angular coordinates of annihilated and created
pairs. Keeping in mind the above remarks, we can write the action for the
electrons in the ring $\pm T_{co}$ around the Fermi surface as
$$
S=\int _0^{\infty}  d\tau \{ \sum _{\sigma}\int _{\epsilon < T_{co}}
\frac{d\epsilon}{2\pi} \oint \frac{ds}{2\pi v( \theta )} \bar{\Psi}
_{\sigma}(\epsilon ,\theta )(\partial _{\tau} -\epsilon )
\Psi _{\sigma}(\epsilon ,\theta ) +
$$
\begin{equation}
+\frac{1}{2}
\sum _{\sigma \sigma'}
\int \frac{d{\bf q}_{ee}}{(2\pi )^2}
\oint \frac{ds}{2\pi v({\theta })}
\oint \frac{ds'}{2\pi v({\theta '})}
\bar{\hat{\Delta}}_{\sigma  ',\sigma ,{\bf q}_{ee}} (\theta ')
V_{l=l_{co}}(\theta ,\theta ')
\hat{\Delta}_{\sigma ',\sigma ,{\bf q}_{ee}}(\theta ) + FL \},
\label{Hlin}
\end{equation}
where $\epsilon =k_{\perp} v(\theta )$, closed loop integrations are over
the the Fermi surface and $FL$ stands for the effective Fermi liquid
interaction.  $\hat{\Delta}_{\sigma ',\sigma ,{\bf q}_{ee}}(\theta )$ is the
energy-integrated number of pairs defined as
\begin{equation} \label{Delta}
\hat{\Delta}_{\sigma _1,\sigma _2, {\bf q}_{ee}}(\theta )\equiv
\int _{\epsilon < T_{co}} \frac
{d\epsilon}{2\pi}\Psi _{\sigma _1}({\bf k})\Psi _{\sigma _2}(-{\bf k}+{\bf
q}_{ee}) \Theta (T_{co} -|(-{\bf k}+{\bf q}_{ee})\cdot \hat{\bf n}v(\theta )|).
\end{equation}
Note that the integration measure over small momentum ${\bf q}_{ee}$ goes to
zero as $l \rightarrow \infty $.  A form similar to (\ref{Hlin}) has been
used by Weinberg\cite{Weinberg}, but taking $V_{l=l_{co}}(\theta ,\theta ')$
phenomenologically and not as the partially renormalized pairing interaction
which we get from $U_{l=l_{co}}(\theta _1,\theta _2,\theta _3)$ putting
incoming particles 1 and 2 to $\theta $ and $\theta +\pi$ and outgoing 3 and
4 to $\theta '$ and $\theta '+\pi $.  Note that the loop integration over
$s$ can be understood as the scalar product over ``vector components'' of a
``spin'', where the number of components $N$ corresponds to $8t/T_{co}$
\cite{Ma,Shankar}.  The integrations have the weight factor $1/v(\theta )$,
$v(\theta )$ being the anisotropic Fermi velocity, what suggests to
introduce a new angular coordinate
\begin{equation}
\label{z}
z(\theta )= \frac{ \int ^s \frac{ds}{v(\theta )}}{2\pi N_F},
\end{equation}
where $N_F$ is the density of states at the Fermi level.  The function
$z(\theta )$ is shown in Fig.4 for few different values of $\mu$.  Starting
from the new Hamiltonian (\ref{Hlin}) we can calculate now the function
$\beta _{ee}\{ V\} $ in $z$-space and obtain the flow-equation

\begin{equation} \label{flowV}
\partial _l { V}(z,z')=-\frac{N_F}{2 \pi }\oint
dz'' { V}(z,z''){ V}(z'',z')\; ,
\end{equation}
where the coordinate $z$ appears instead of $\theta (z)$.  For initial
condition we take $V_{l=l_{co}}(\theta (z),\theta (z'))$.

\section{Diagonalization of the pairing potential}
To make the differential equation (\ref{flowV}) solvable one has to
diagonalize the pairing potential ${V}(z,z').$\cite{Shankar,Weinberg} Since
it is invariant under all symmetry elements of the $D_4$ point group, its
most general form in $z$-space can be written as
\begin{equation} \label{decomp}
V(z,z')=\sum _\gamma \sum _{m,n}
V_{m,n}^{\gamma}f_{m,n}^{\gamma}(z,z')\; ,
\end{equation}
where $V_{mn}^{\gamma}\equiv \langle m\gamma |V|n\gamma \rangle $ and
$f_{m,n}^{\gamma}(z,z')\equiv \langle m\gamma |z\rangle \langle n\gamma |z'
\rangle $.  The function $\langle m\gamma |z\rangle$ is the $m^{th}$ basis
state of the $\gamma$--representation of the point group $D_4$. It is
proportional to the function $\cos 4mz$, $\sin 4mz$, $\cos (4m+2)z$, $\sin
(4m+2)z$, and $[\cos (2m+1)z \pm \sin (2m+1)z]$, for $\gamma =A_1 ,\; A_2
,\; B_1 ,\; B_2$, and $E$ respectively.  Using (\ref{decomp}), the flow
equation (\ref{flowV}) becomes
\begin{equation} \label{flowdec}
\partial _l  V_{m,n}^{\gamma}=-\frac{N_F}{2\pi }\sum _{\nu}
V_{m,\nu}^{\gamma}V_{\nu,n}^{\gamma}
\end{equation}
with the initial condition
\begin{equation} \label{initdec}
V_{m,n}^{\gamma}(l=l_{co})=\int dzdz'
f_{m,n}^{\gamma}(z,z') V_{l_{co}}(z,z').
\end{equation}
To solve exactly the equation (\ref{flowdec}) one has to diagonalize five
infinite dimensional matrices $V_{m,n}^{\gamma}(l=l_{co})$, thus decoupling
completely the flow (\ref{flowdec}) into a set of differential equations
whose solution is
\begin{equation} \label{solution}
 V^{\gamma}_{\lambda}(l)=\frac{V^{\gamma}_{\lambda}(l_{co})}
{1+(\frac{N_F V^{\gamma}_{\lambda}(l_{co})}{2\pi
})(l-l_{co})}.
\end{equation}
Here $\lambda$ labels the eigenvalues within the representation $\gamma$.
If $V^{\gamma}_{\lambda}(l_{co})$ is negative, the denominator has a zero at
$l=l_c(\gamma ,\lambda )$ and an instability occurs.

The renormalization group calculation of the pairing interaction $V(l_{co})$
is an extremely difficult problem because of the interplay between the
cutoff $\Lambda$ and the geometry of the two-dimensional shells
$d\Lambda$. However for some range of $\langle n\rangle$ and $U_0$
(essentially small $U_0$ and $\langle n \rangle$ not too close to
half-filling) the problem can be reduced to only one simple integration,
i.e.
\begin{equation} \label{Vco}
V(l_{co})\approx U_0+I_0
\end{equation}
with
\begin{equation}  \label{I0}
I_0=U_0^2(P_{ee}(l_{co})+P_{eh}(l_{co})) ,
\end{equation}
where $P_{ee}(l)$ is the partially integrated e-e loop, defined in a same
way as $P_{eh}(l)$ in eq.(\ref{Bubble}). $P_{eh}(l)$ comes just from the last
diagram in the expression for $\Gamma _4$ in Fig. 1, because the other three
e-h loops cancel exactly for any $U$ symmetric under exchange. The
approximation (\ref{Vco}) is allowed if $I_0/U_0$ is a small parameter.  A
good estimate of the magnitude of this parameter is given by the $A_1$-part
of $U_0P_{eh}(l_{co})$, giving the criterion for the validity of our
approach in $(\langle n\rangle ,U_0)$-space. The line $I_0(\mu )/U_0\sim 1$
is shown in Fig. 5. Below the line the approximation (\ref{I0}) is
justified.

By definition of $l_{co}$, $P_{eh}(l_{co})\approx P_{eh}(l=\infty
)$. Moreover, $P_{ee}(l_{co})$ has no dependence on $z$ and $z'$, since it
depends on external momenta only through ${\bf k_1 +k_2}$, which we put to
zero. Consequently, its only nonzero component is $\langle
0A_1|P_{ee}|0A_1\rangle $. Thus, for the calculation of all other components
of $V_{l_{co}}$ we use just the bubble $P_{eh}(l\rightarrow \infty)$ with
the momentum transfer ${\bf q}={\bf k}_F(z)-{\bf k}_F(z'+\pi )$. Fig.6(a)
shows $P_{eh}(l\rightarrow \infty)$ as a function of $z$ and $z'$ for
chemical potential $\mu /4t=-0.2$.

The minimal eigenvalues of $V(l_{co})$ in all five channels, named
$\lambda_{\min}^{\gamma}$, are shown in Fig.6(b) as functions of
$\mu$. These curves indicate which kind of superconducting symmetry becomes
critical at some given $\mu$. The eigenvalues for each channel are
calculated taking only the first four harmonics for $A_1$, $A_2$, $B_1$ and
$B_2$, and the first six harmonics for the $E$ representation.  The
corresponding eigenvectors determine the Fourier spectrum of the gap
function. A very important result is that the relevant harmonic of the
superconducting fluctuations in the $B_1$ channel occurs very close to just
$\cos (2z)$, being thus determined only by the structure of the Fermi
surface and not by the interaction.  Fig. 7(a) show the instable order
parameters $\Delta _{B_2}$ for $\mu =-0.5$, $\Delta _E$ for $\mu =-0.31$ and
$\Delta _{B_1}$ for $\mu =-0.001$ as a function of the Fermi surface angle
$\theta $. The evolution of the function $\cos 2z(\theta )$ (i.e. the first
harmonic of $B_1$) with $\log (-\mu )$, given only by the dependence of $z$
on $\theta$, is shown in Fig. 7(b). The strength of the peaks near the van
Hove points increases and the magnitude in the area between the peaks
decreases with $\log (-\mu )$.

The critical temperature is given by a cutoff for which the most attractive
diagonal component of $V$ diverges, i.e.
\begin{equation} \label{Tc}
T_c = 8t \exp [-l_c(\gamma ,\lambda _{\min})],
\end{equation}
where $\lambda _{\min} = {\min} \{ \lambda _{\min}^{\gamma}\} $.
Fig.8. shows $l_c$ as a function of $\log (\mu )$. The critical temperature
decreases extremely fast as we go away from the half-filling.  An increase
of $U_0$ could save the situation, but in that case our perturbative method
ceases to be sufficient (see Fig. 5). Since the cuprates are superconductors
for fillings quite far from one electron per site ($\langle n \rangle \sim
1-0.17$), this result means that the small-$U$ Hubbard model cannot describe
these systems quantitatively. However, the model gives very precious
informations about the form of the gap function in the $B_1$--instable
regime, which will not change considerably with increasing $U_0$, as long as
$\cos (2z(\theta ))$ is the dominant attractive harmonic in $V(l_{co})$.

\section{Mixed--symmetry superconductivity}
Once the renormalization flow has been integrated for $l<l_{co}$, assuming
that the interaction did not diverge earlier in the antiferromagnetic
channel, the detailed angular dependence of the superconducting gap function
can be easily found. In general, a superconducting state with the symmetry
corresponding to the lowest of the eigenvalues $\lambda^\gamma_{\rm min}$
will be formed. However, when two of the $\lambda^\gamma_{\rm min}$ are
close to each other, a more complicated situation can occur: for
definiteness, consider the region $0.206<|\mu /4t|<0.276$ in Fig.6(b), where
the $B_2$ eigenvalue is the most attractive after the $B_1$.
Let us suppose that $B_1$
order of the simplest form $\Delta _{B_1}\sim \cos 2z$ has formed and that
the temperature is close to $T_c$. Among the remaining symmetry channels,
$B_2$ is the only one which can give a large gap function in the node-points
of $\Delta _{B_1}$ and zero in the points where $\Delta _{B_1}$ is maximal.
Consequently, we expect that the flow of the type (\ref{solution}) with
$\gamma =B_2$ will not be strongly affected by the existing $B_1$
order. Considering to a first approximation the two flow equations ($\gamma
=B_1$ and $\gamma =B_2$) as independent, and taking only the first harmonics
of the $B_1$ and $B_2$ representations, we can construct the relevant part
of the pairing interaction which gives two phase transitions, one with $B_1$
and the other with $B_2$-symmetry:
\begin{equation} \label{Vapr}
V_{l_{co}}(\theta _1,\theta _2)=V^{(B_1)} \frac{1}{\pi}\cos 2z(\theta _1)
\cos 2z(\theta _2)+V^{(B_2)}
\frac{1}{\pi} \sin 2z(\theta _1) \sin 2z(\theta _2) ,
\end{equation}
where all details of the Fermi surface are contained in the dependence of
$z$ on $\theta$.  From eq.(\ref{solution}) one finds that the ratio between
two critical temperatures is given by $T_c'/T_c=\exp [-2\pi
(1/V^{(B_1)}-1/V^{(B_2)})/N_F]$. From Fig.6(b) note that the ratio
$T_c'/T_c$ is very sensitive to the variation of the chemical potential.
The gap function resulting from (\ref{Vapr}) has the form
\begin{equation} \label{Delta1}
\Delta (\theta )= \Delta _{B_1}e^ {i\phi _1} \cos 2z(\theta )
+\Delta _{B_2}e^ {i\phi _2} \sin 2z(\theta )\; ,
\end{equation}
where $\Delta _{B_1}$, $\Delta _{B_2}$, $\phi _1$ and $\phi _2$ are
real. These parameters can be determined minimizing the mean-field
expression for the free energy per site\cite{Kotliar}
\begin{equation} \label{Flandau}
F=-\frac{2T}{N}\sum _{\bf k} \Theta(T_{co}-|\xi _{\bf k}|) \ln \cosh
\frac{E_{\bf
k}}{2T}+ |\Delta _1|^2/V^{(B_1)}+|\Delta _2|^2/V^{(B_2)},
\end{equation}
where $E_{\bf k}\equiv \sqrt{\xi _{\bf k}^2+|\Delta (\theta )|^2}$ and the
theta function constrains the momentum summation to run only over the states
within the energy shell $\pm T_{co}$ about the Fermi surface. The
minimization of $F$ with respect to $\cos (\phi _1 -\phi _2)$ gives
\begin{equation} \label{phi}
\phi=\phi _1 -\phi _2=\pm \frac{\pi}{2},
\end{equation}
i.e. the resulting gap function is of the type $B_1\pm iB_2$. It is
interesting to remark that the same kind of gap function has been obtained
by Laughlin using the anyon picture.\cite{Laughlin} The particularity of
this gap function (and of any gap consisting of two different symmetry terms
with a phase difference of $\pm \pi/2$) is that it breaks time reversal
symmetry.

We can now try to understand recent ARPES measurements by Jian Ma and
coworkers \cite{JianMa} on the Bi2212 compound. From their experiment it
appears that two superconducting instabilities occur; the first one is at
$T=T_c$ and has probably the $B_1$ symmetry. The second instability occurs
at $T_c'=0.81T_c$; it introduces a nonzero gap at the points $\theta
=(2n+1)\pi /4$, i.e. halfway between the corners of the half-filled
Fermi. The function measured by ARPES is $|\Delta (\theta )|$ and in the
picture discussed above has no zeros and minima on the diagonals of the
Brillouin zone if both $\Delta _{B_1}$ and $\Delta _{B_2}$ are finite,
$|\Delta _{B_1}|>|\Delta _{B_2}|$ and $\phi =\pm \pi /2$. This is in
agreement with the experiments because the gap in the diagonal direction is
just equal to $\Delta _{B_2}$, introduced at $T=T_c'$. The minimum of
$|\Delta (\theta )|$ on the diagonals is in agreement with other ARPES
experiments\cite{Shen,Norman} as well. One should of course notice that in
our model closeness of two different $\lambda$'s only occurs in very narrow
parts of the parameter space and therefore to a certain degree is accidental.

\section{Conclusions}
We have formulated the one-loop renormalization group for a two-dimensional
system of interacting electrons on a square lattice, described by the
Hubbard model. For band filling different from one-half, the renormalization
flow for $l$ superior to some crossover $l_{co}$ comes only from the
contribution due to the electron-electron bubble-diagram, while the
electron-hole contribution decays exponentially as $\exp(-l/2)$,
where $l_{co}$ depends on band filling, but is independent on the strength
of the interaction.  We decompose the BCS pairing interaction $V$ for
electrons in the vicinity of the Fermi surface in Fourier components of the
five irreducible representations of the $D_4$ point group, defined at the
Fermi surface. Diagonalizing $V(l_{co})$ in each representation, we get five
sets of decoupled BCS flow equations. The minimal (i.e. the most negative)
eigenvalue of $V(l_{co})$ determines the critical temperature and the
eigenvector gives the form of the gap function. Unlike the usual
approach,\cite{Gorkov} the characteristic of the procedure presented here to
obtain the symmetry of the gap function is that only the axial coordinates
at the Fermi surface is relevant, while the radial dependence is ``scaled
out''. Moreover, the renormalization group treatment of the whole Brillouin
zone, and not only of the narrow belt about the Fermi surface has allowed us
to show that the origin of the attractive part of the pairing interaction in
the Hubbard model is in electron-hole fluctuations on rather high energy
scales, up to the bandwidth.

We have calculated $V(l_{co})$ for the case where the flow due to the
electron-hole channel can be treated perturbatively, i.e. when the filling
is far enough from one-half. The diagonalization of $V$ in terms of angular
harmonics gave us the type of superconducting instability: for weak filling,
the instability occurs in the $B_2$ ($d_{xy}$) singlet channel, while for
filling close to one-half, the $B_1$ ($d_{x^2-y^2}$) singlet instability
strongly overwhelms all others, what are the results in agreement with previous
work.\cite{Baranov,Scal2,FLEX1} Particularly interesting is the fact that
the order parameter $\Delta _{B_1}(\theta )$ can be very well approximated
by the function $\cos 2z(\theta )$, where the function $z(\theta )$ depends
only on the anisotropic Fermi velocity and on the geometry of the Fermi
surface.  This gives for $\Delta _{B_1}(\theta )$ a function that has peaks
in the directions of the van Hove singularities.  The slope of the peaks
increases as we approach half-filling.  This can be a justification to
consider the interaction only between electrons in the close vicinity of the
van Hove points as relevant if we are very close to half-filling
\cite{HJS}. We believe that the form of $\Delta _{B_1}(\theta )$ does not
depend considerably on the strength of interaction, and recent T-matrix
calculations \cite{Dahm} for realistic values of the interaction $U_0$ give
a gap function in accord with our assumptions.

We find a  superconducting instability at
any
 electron
concentration away from half--filling.
 The underlying physical mechanism, namely
exchange of spin or charge density fluctuations, is the same as in
previous approaches.\cite{Emery,Miyake,Scal1,Scal2,FLEX1,FLEX2,FLEX3}
 We do however feel that our present results are on a
more solid footing than the previous work because the present one--loop
renormalization group scheme does not make any a priori assumptions about
important or unimportant diagrams and provides a more systematic way of
handling the dynamics of the fluctuations being exchanged. The only
restrictions come from (i) the limitation to one--loop order, necessitating
weak coupling, and (ii) the requirement that the e-h diagram are a
perturbation with respect to the e-e diagrams, implying that we can not be
too close to half--filling. The region of validity of the approach is shown
in Fig.5. Further, self--energy diagrams have been neglected, however, these
are expected to produce important effects only at two--loop order, and
therefore are expected to be negligible in weak coupling.

In our weak coupling model the superconducting critical temperature is
negligible (but it exists, for any filling!) if we are not in the immediate
vicinity of half-filling, which means that the Hubbard model with small
$U_0$ and small (perturbative) antiferromagnetic fluctuations does
not suffice to describe the high critical temperature ($\sim 0.02t$) of the
cuprates. There exist two possible ways (related to two restrictions of our
calculations) to increase the critical temperature. The first is to simply
increase $U_0$ and to remain far from the half-filling, keeping the e-h
channel non-singular. To treat this case, an approach perturbative in $U_0$
is possibly only of limited use.  Ideally, renormalization should be done
exactly, and not using a simple one-loop (or $n$-loop) scheme (which is
actually just an ``intelligent'' version of the perturbative summation). We
can speculate and suppose that even in the case of strong coupling there
exists the crossover $\tilde l_{co}$, above which the flow is of the BCS
type. It is to expect that $\tilde l_{co}$ is not very different from
$l_{co}$ that we have calculated.  This means that the cutoff $\tilde
T_{co}$ for the effective BCS theory is $\Lambda _0\exp (-l_{co})$ (see
Fig.3. and eq.(\ref{co})), where $\Lambda _0$ is the initial cutoff of the
theory, equal to $8t$.  The second possibility to increase $T_c$ is to
approach half-filling very closely, making Fermi surface nesting important
but remaining in a weak-coupling regime. For that case, a simplified
one-loop renormalization group calculations \cite{HJS} has shown that
superconductivity wins over antiferromagnetism only if the e-h contribution
to the flow decays before the divergence in the antiferromagnetic channel
takes place. Thus, we can say that here too, the effective theory is of BCS
type. The difference with the first scenario is that the effective cutoff is
very small (Fig. 3), and that the coupling constant is very strong, due to
the strong flow in both e-e and e-h channels at all scales $l<l_{co}$. A
very important feature of the nested case with a small $U_0$ is that it can
be treated in terms of the one-loop renormalization group, renormalizing $U$
as a function only of three angular variables. This is allowed because all
important physics (i.e. the majority of the e-e and e-h flow) is contained
in the vicinity of the Fermi surface, making the effective phase space to be
a rather narrow square ring $\pm \Lambda _i$; $(8t\gg \Lambda _i \gg \Lambda
_{co})$ where the marginally relevant interaction is a function only of the
angular position of the particles on the ring.

Finally, we have discussed the possibility of a superconducting state with a
mixed symmetry.  In the presence of $B_1$ order, the flow in the $B_2$
channel (which is the second most attractive one for $0.206< \mu <0.276$)
will be only weakly affected by a nonzero order parameter of $B_1$
symmetry. This gives rise to two superconducting instabilities, with the
critical temperatures $T_c$ for the $B_1$ and $T_c'$ for the $B_2$ channel,
and $T_c'<T_c$. We have given the form of the pairing function for the
effective BCS theory. At $T<T_c'$ the relative phase of two order parameters
is $\phi =\pm \pi /2$. The resulting form of the energy gap $|\Delta
(\theta)|$ has no zeros and minima are in diagonal directions, providing a
possible qualitative explanation of ARPES experiments by Shen\cite{Shen} and
the decrease of the anisotropy with decreasing $T$.\cite{JianMa}

\acknowledgements

We are indebted to N. Dupuis and L. Guerrin for a number of helpful
discussions. This work was supported by EC contract no. ERBCHRXCT 940438.

\newpage
\begin{center}
{\bf FIGURES}
\end{center}

FIG. 1. The one-particle irreducible diagrams for the vertices $\Gamma _2$ and
$\Gamma _4$,  generating the renormalization of the self-energy and
the interaction, respectively.

FIG. 2. The e-h bubble (solid line) and its derivative over $l$ (dashed
line) for
$\mu /4t=-0.25$ (a) and $\mu /4t=-0.02$ (b), for momentum transfer ${\bf
q}_{eh}=2{\bf k}_F\parallel (\pi ,\pi )$.

FIG. 3. The scale $l_x$ as a function of filling $\langle n \rangle$. For
$l>l_x$ the e-h flow decays exponentially. The insert shows the relation
between  $\mu$ and $\langle n \rangle$.

FIG. 4. The relation between the angular variable $z$ and the observable polar
angle $\theta$ for $-\mu /4t=2 \times 10^{-n}; \; n=1(a),2,...,9(i)$.

FIG. 5. The curve $U_0P_{eh}(l\rightarrow \infty )=1$. Below the curve, the
e-h contribution to the renormalization can be treated perturbatively.

FIG. 6. (a) The shape of the function $P_{eh}(l\rightarrow \infty )$ in
$(z,z')$ space at $-\mu /4t=0.2$. The nesting at half filling
occurs for $z=z'= \pi
/4$. The split singular lines show the best incommensurate nesting
vector. (b) The minimal eigenvalue of the pairing $V(z,z')$ in every of 5
irreducible representations of $D_4$ point group.

FIG. 7. (a) The shape of three possible gap functions: $\Delta
_{B_1}(\theta)$ for $\mu /4t=-0.001$ (dot-dashed); $\Delta _{B_2}(\theta)$
for $\mu /4t=-0.5$ (dashed) and $\Delta _{E}(\theta)$ for $\mu /4t=-0.31$
(solid line).  (b) A very good approximation for $\Delta _{B_1}(\theta)$ is
just $\cos 2z(\theta )$, shown here for the same choice of $\mu$ as in
fig. 4. The dashed line shows $\cos \theta$ to comparison.

FIG. 8. The scale $l_c=-\ln T_c/8t$ as a function of the logarithm of the
chemical potential, for a few values of coupling $U_0$. For very small $\mu$
and for $U_0/4t > 0.5$ the curves are out of the range of validity (see
fig.5).

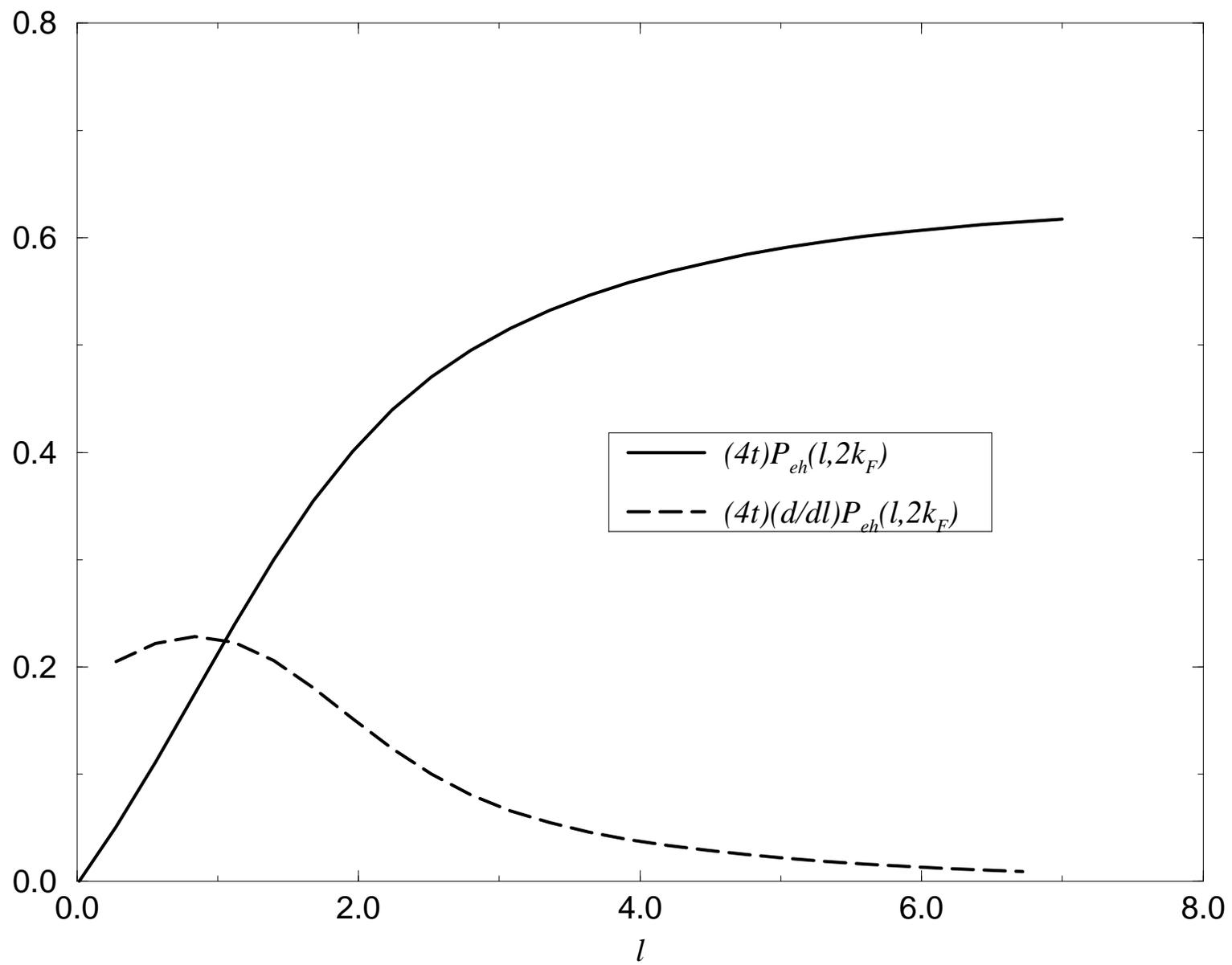
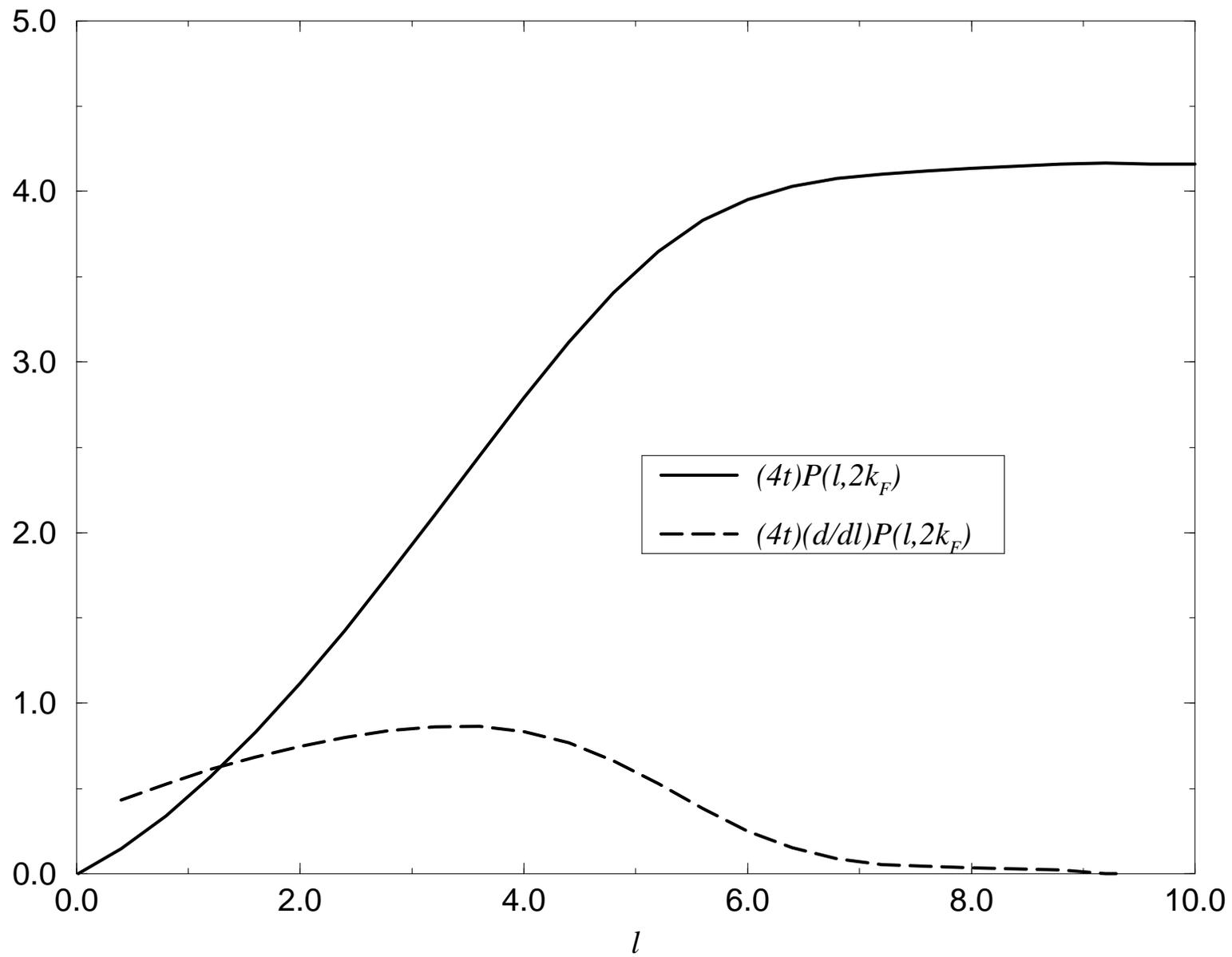
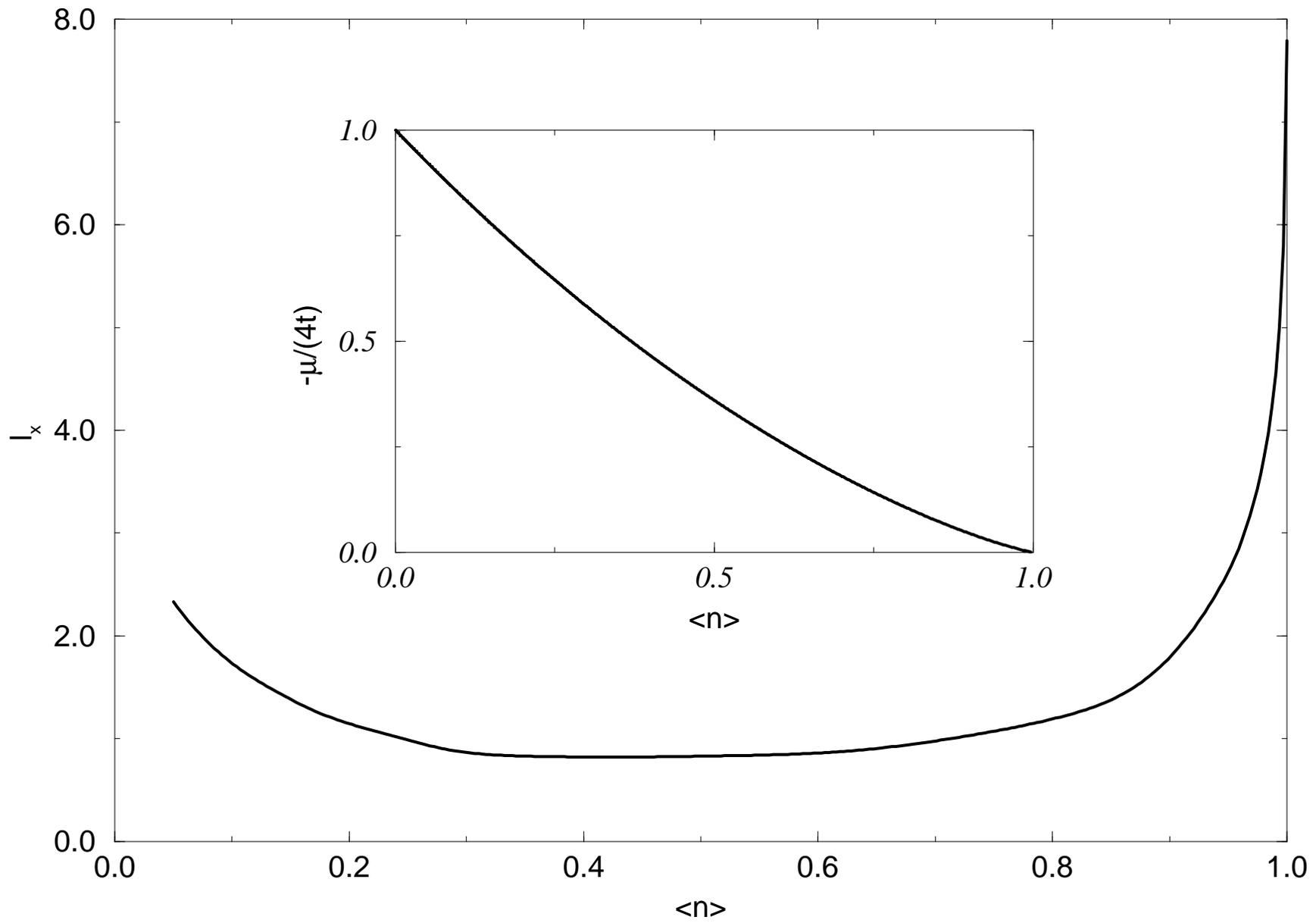
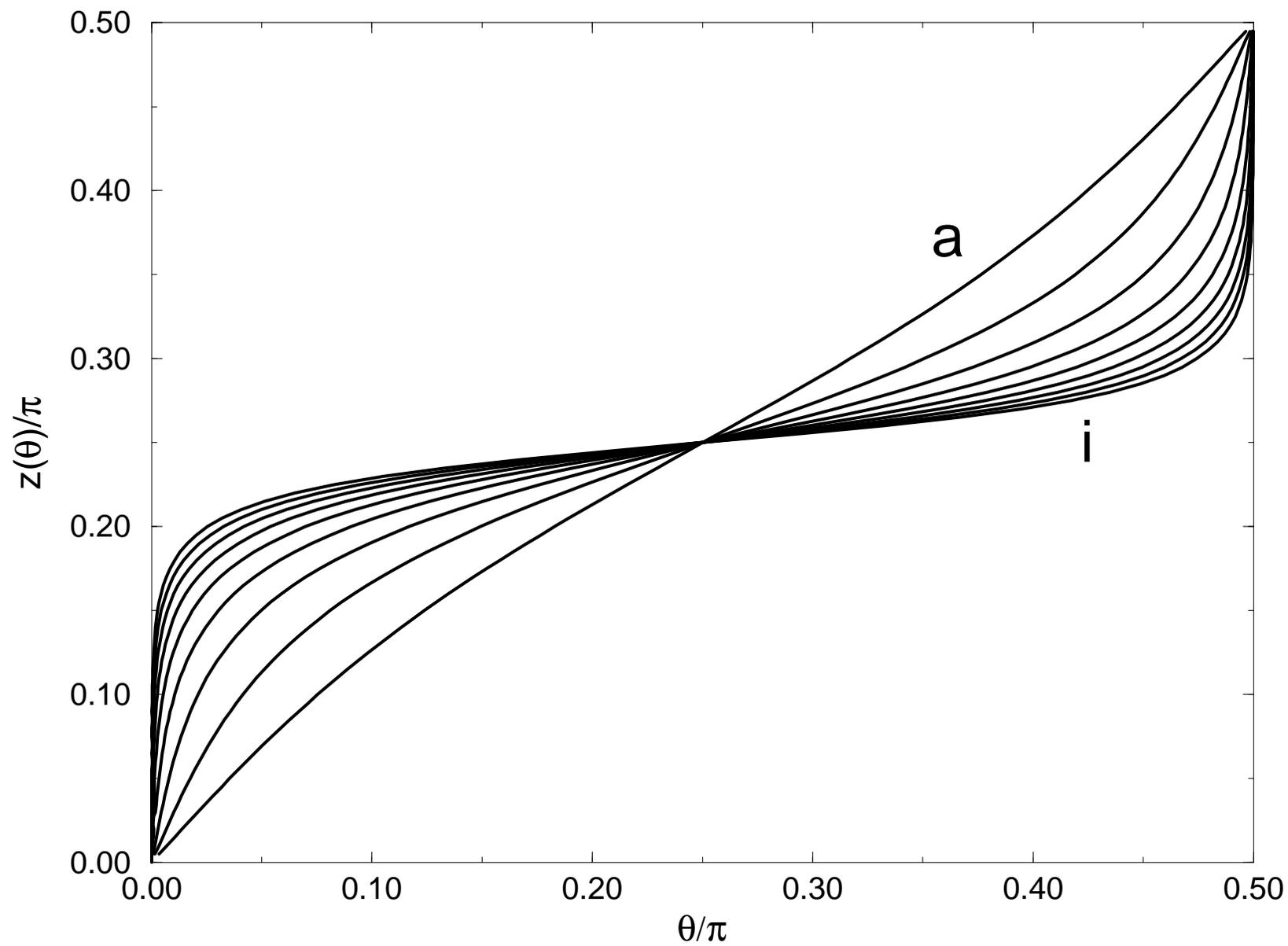
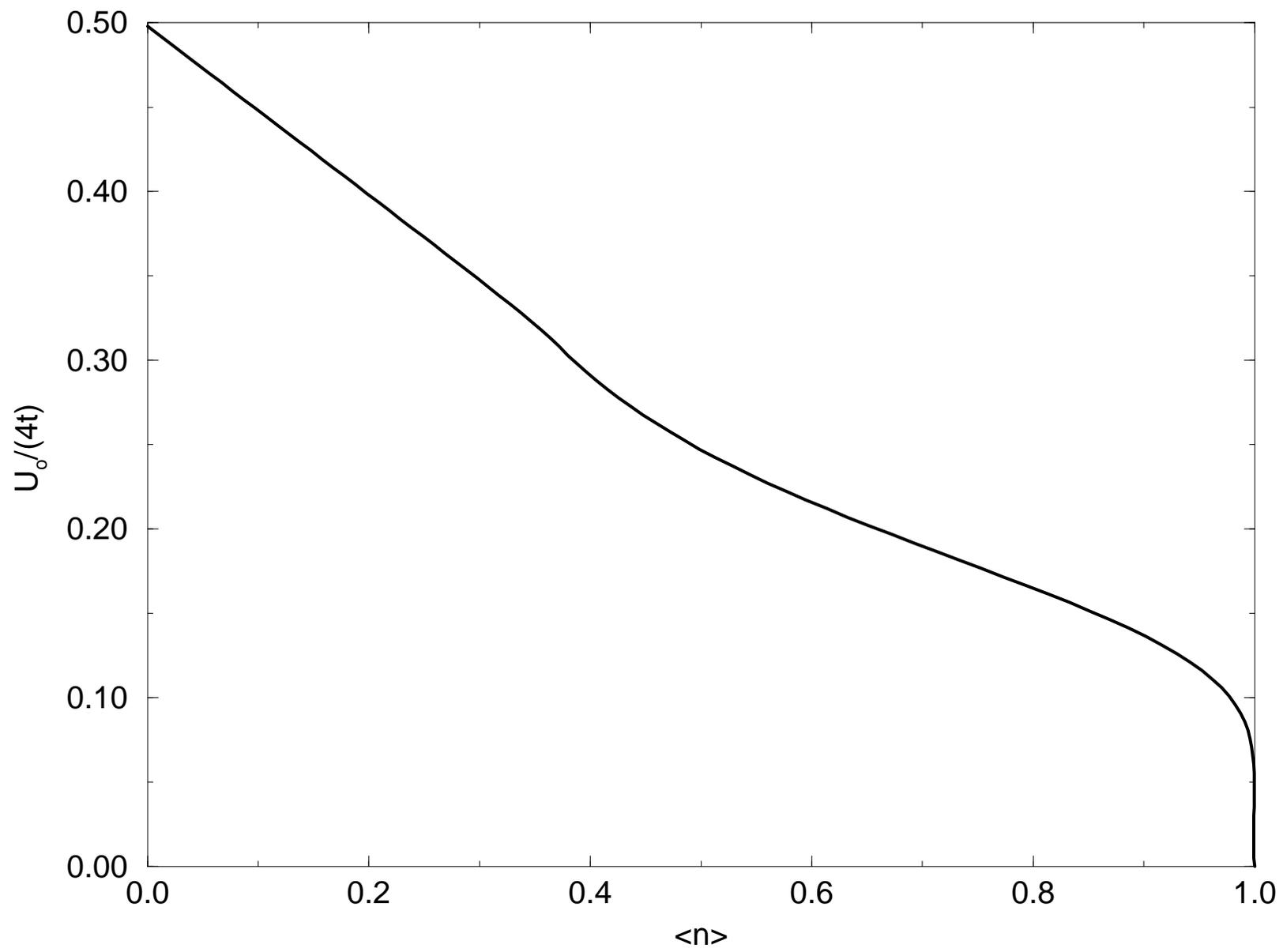
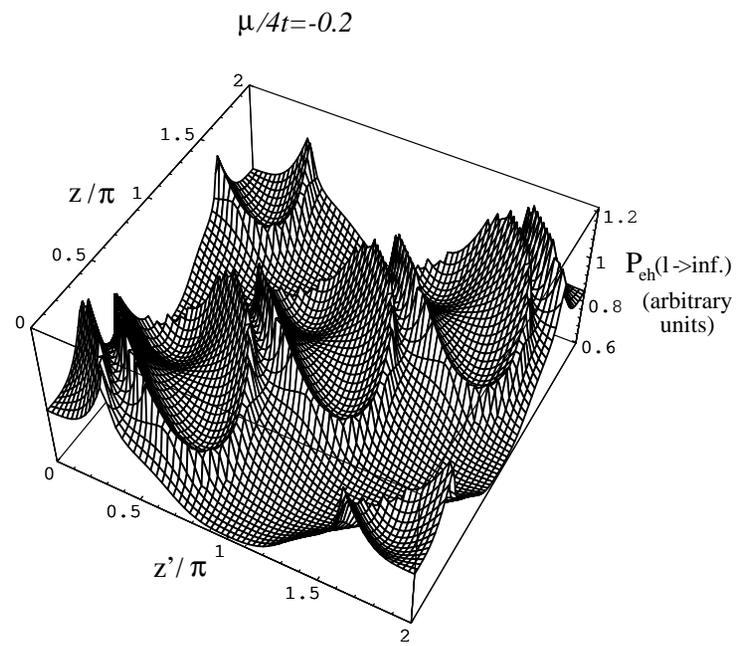
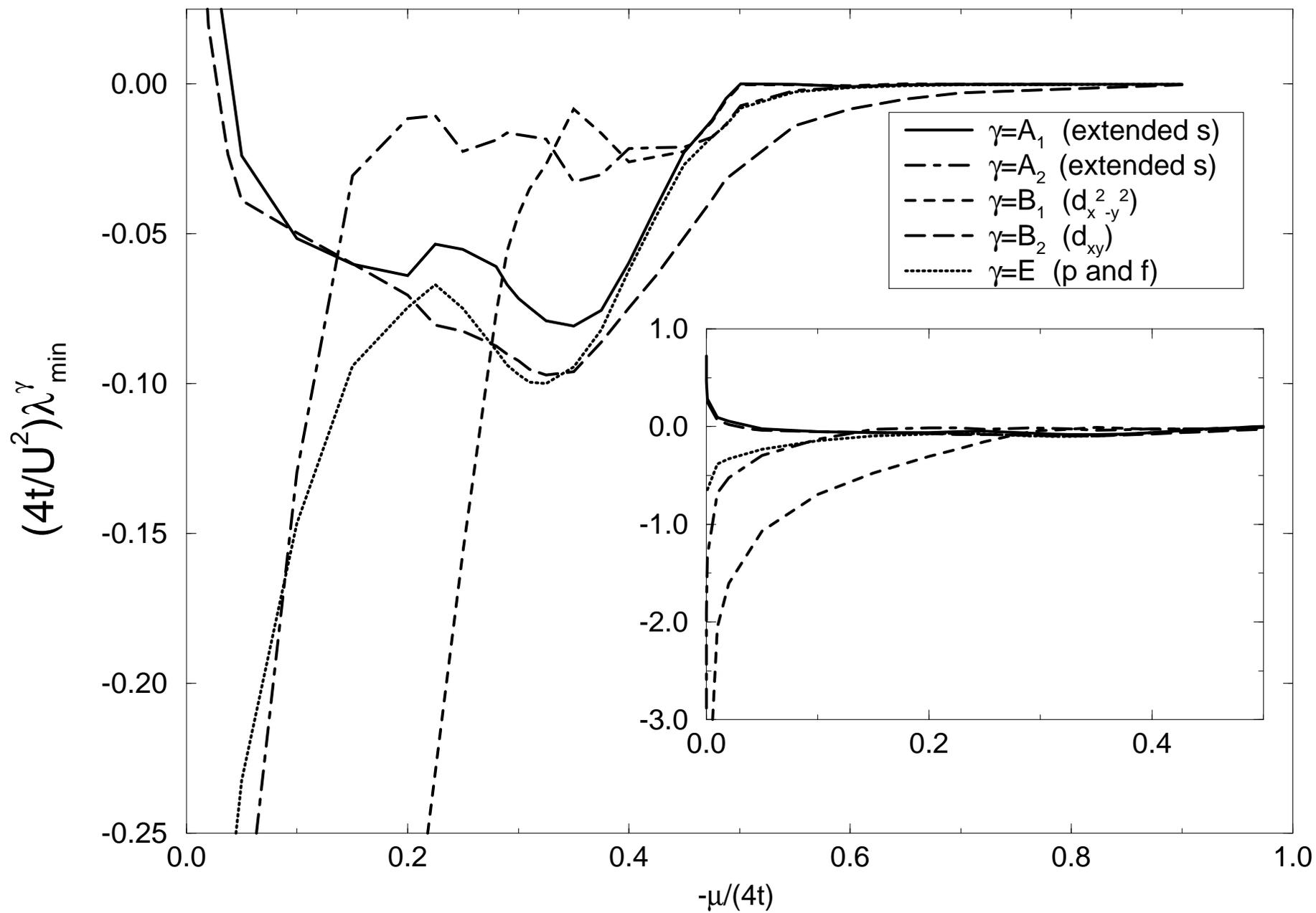
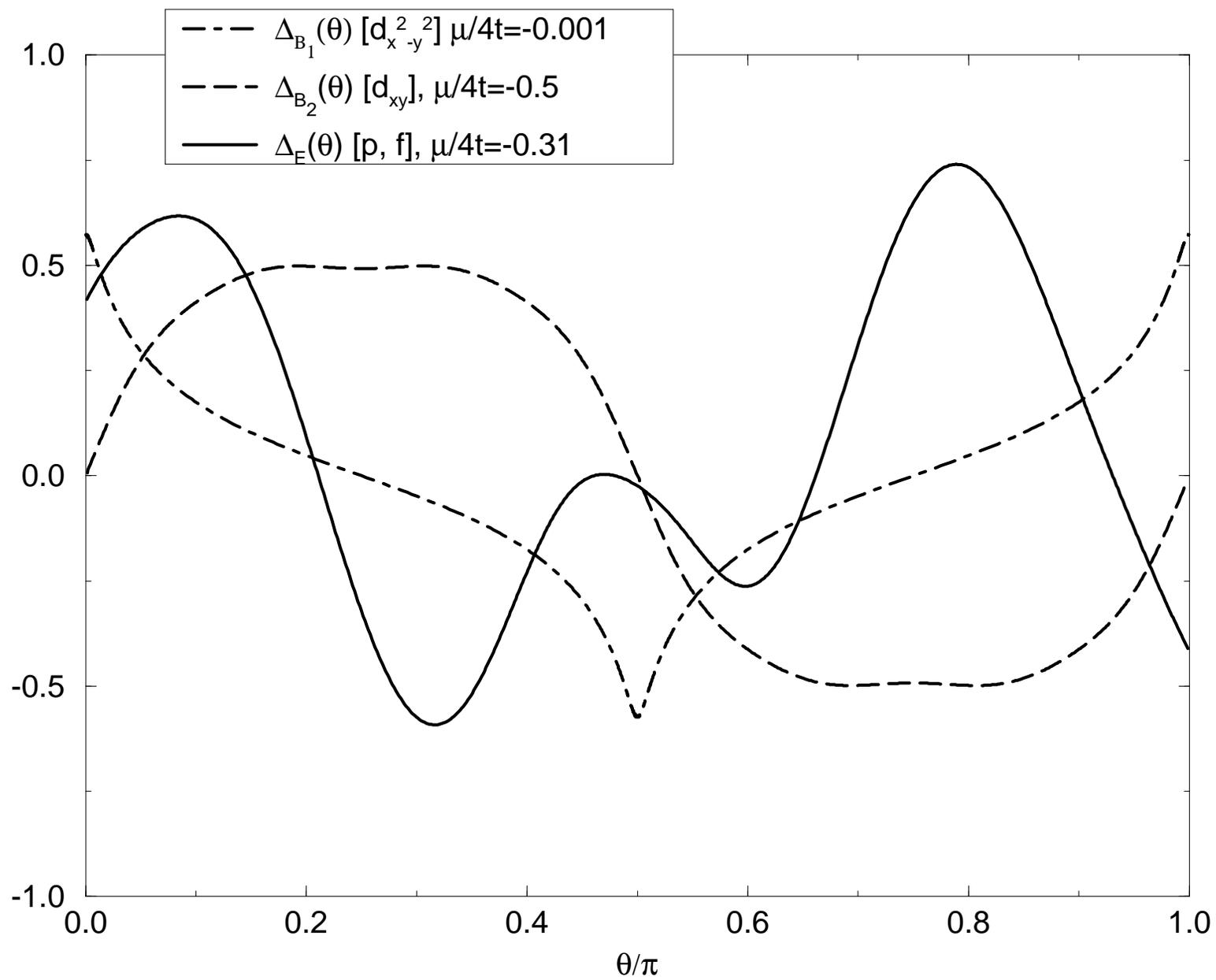
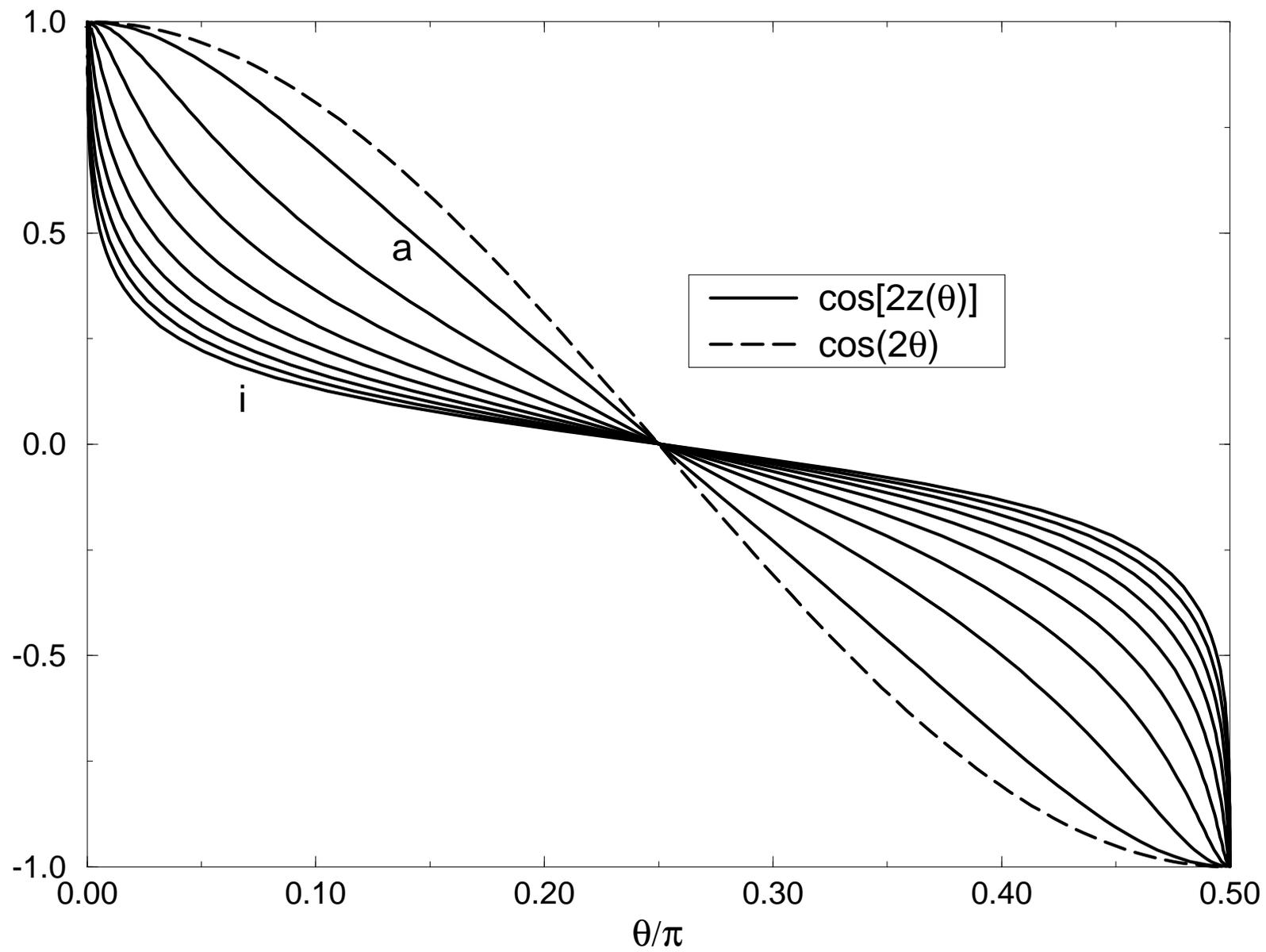
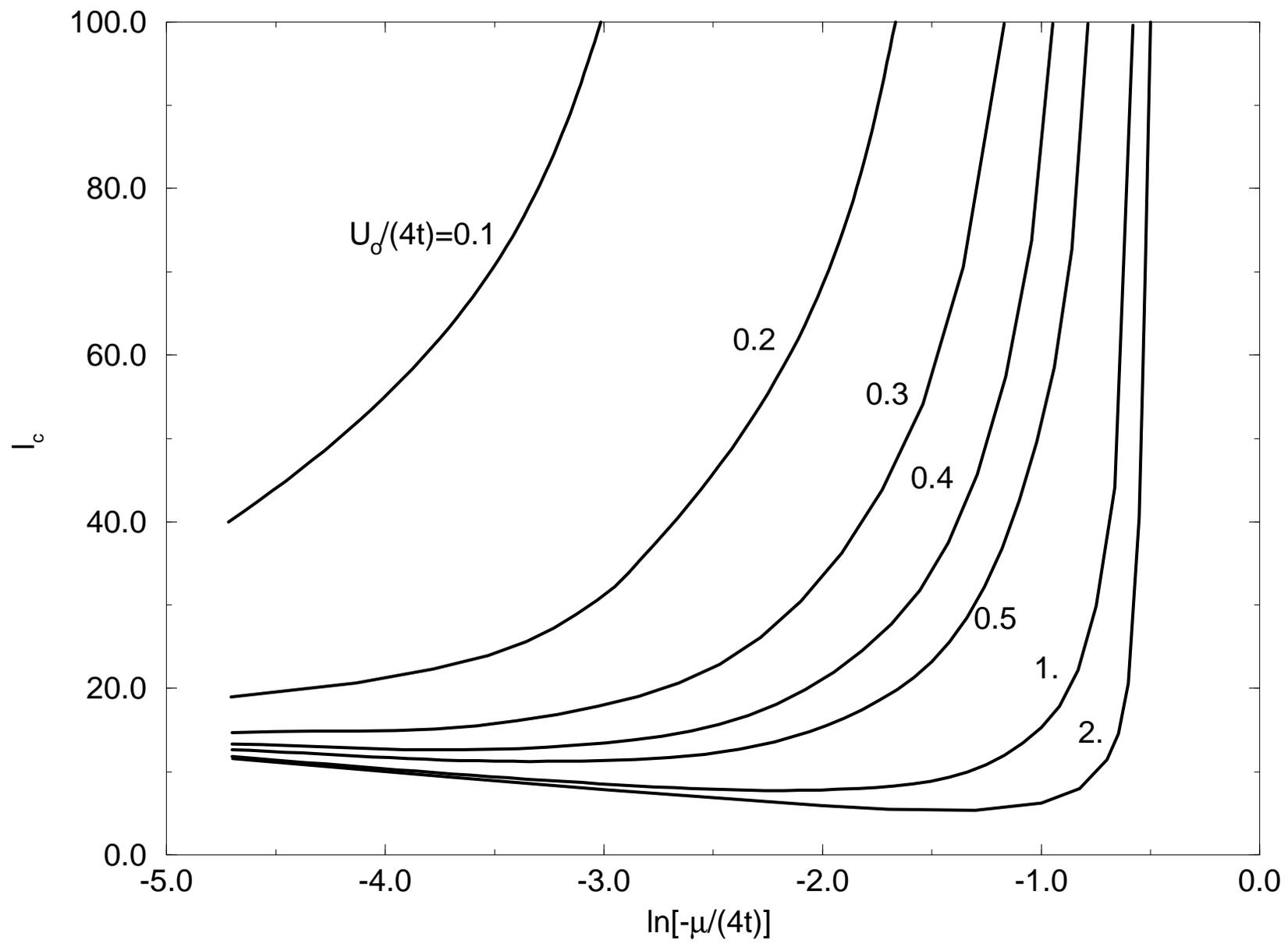
\end{document}